\documentstyle[prl,aps,preprint]{revtex}
\input epsf
\tightenlines
\newcommand{\ppi}{\overline{\psi}}
\newcommand{\g}{\overline{g}}
\newcommand{\gphys}{\tilde{g}}
\newcommand{\oli}[1]{\overline{#1}}

\newcommand{\half}{{1\over2}}

\newcommand{\gcal}{{\cal G}}
\newcommand{\fcal}{{\cal F}}

\newcommand{\Vbar}{\oli{V}}

\newcommand{\spatial}{\int^{\sigma_+}_{\sigma_-}dx}

\newcommand{\ba}{\begin{array}}
\newcommand{\ea}{\end{array}}

\newcommand{\be}{\begin{equation}}
\newcommand{\ee}{\end{equation}}
\newcommand{\bea}{\begin{eqnarray}}
\newcommand{\eea}{\end{eqnarray}}
\newcommand{\beal}{\setcounter{letter}{1} \begin{eqnarray}}
\newcommand{\eeal}{\addtocounter{equation}{1} \end{eqnarray}}

\newcommand{\req}[1]{Eq.(\ref{#1})}

\newcommand{\larrow}{\,\,\,\,\hbox to 30pt{\rightarrowfill}
\,\,\,\,}
\newcommand{\slarrow}{\,\,\,\hbox to 20pt{\rightarrowfill}
\,\,\,}
\begin{document}
\setcounter{page}{0}
\def\footnoterule{\kern-3pt \hrule width\hsize \kern3pt}
\tighten
\title{Hamiltonian Thermodynamics of Black Holes in Generic\\
    2-D Dilaton Gravity}

\author{G. Kunstatter\footnote{Email address: {\tt gabor@theory.uwinnipeg.ca}}\\
    R. Petryk\\
     S. Shelemy}

\address{Winnipeg Institute for Theoretical Physics \\
and \\
Physics Department, University of Winnipeg \\
Winnipeg, Man., Canada R3B 2E9\\[5pt]
{~}}

\date{hep-th/9709043. {~~~~~} September 1997}
\maketitle

\thispagestyle{empty}

\begin{abstract}
We consider the Hamiltonian mechanics and thermodynamics of 
an eternal black hole in a box of fixed radius and temperature
in generic 2-D dilaton gravity. Imposing boundary
conditions analoguous to those
 used by  Louko and Whiting for spherically 
symmetric gravity, we find that 
the reduced Hamiltonian generically takes the form:
$$
H(M,\phi_+) = \sigma_0 E(M,\phi_+) -{ N_0\over 2\pi} S(M)
$$
where $E(M,\phi_+)$ is the quasilocal energy of a black hole of mass $M$
 inside a static
box (surface of fixed dilaton field $\phi_+$) and $S(M)$ is the associated 
classical
thermodynamical entropy. $\sigma_0$ and $N_0$ determine time evolution along
 the world line of the box and boosts at
the bifurcation point, respectively. An ansatz for the 
 quantum partition function is
obtained by fixing $\sigma_0$ and $N_0$ and then tracing the operator 
$e^{-\beta H}$ over mass eigenstates. We
analyze this partition function
 in some detail both generically and for the class
of dilaton gravity theories that is  obtained by dimensional reduction of
Einstein gravity in n+2 dimensions with $S^n$ spherical symmetry. 
\end{abstract}

\vspace*{\fill}
\begin{center}
Submitted to: {\it Physical Review D}
\end{center}

\pacs{xxxxxx}
\clearpage

\section{Introduction}\medskip
\par
Although it has been
more than twenty years since its discovery, the  thermodynamic behaviour
 of black holes\cite{hawking,bekenstein1} is still not well
understood.
One of the most compelling mysteries is the microscopic
source of the  Bekenstein-Hawking
 entropy.
Some progress has been made recently in specific
contexts. String theory\cite{stringy} has successfully accounted for the
microscopic states for certain
extremal and near extremal black holes. 
A completely different and more geometrical approach has 
been proposed by Carlip\cite{carlip} and by Balachandran and collaborators
\cite{balachandran}. According to this proposal
 black hole boundary conditions give rise
to surface terms at the 
 horizon that  break the diffeomorphism invariance
of the theory. Diffeomorphism invariance can only
 be restored by assuming the existence of
new, physical modes that live on the horizon\footnote{An alternative, and
effectively equivalent description states that as a result
of the black hole boundary conditions, some diffeomorphism modes on the
horizon become physical.}.  Carlip\cite{carlip}
 has counted the resulting edge states in the gauge 
theory formulation of 2+1 anti-deSitter gravity to obtain 
 the correct entropy of BTZ black holes\cite{btz}\footnote{This proposal
however seems to need some modification when applied to dilaton gravity
theories in 1+1 spacetime dimensions\cite{gks}.}.  Another interesting proposal, due to Jacobson\cite{jacobson},
invokes Sakharov's theory of induced gravity\cite{sakharov}. In this approach
 gravity emerges as a macroscopic, bulk theory
obtained by integrating out quantum fields in the 
effective action. Frolov, Fursaev and Zelnikov
\cite{frolov} have shown that induced gravity can in at least some cases
 successfully account
for black hole entropy in terms of microscopic states of the
underlying  quantized fields. 
\par
The fact that these very diverse approaches can all be made to work,
suggests that the correct explanation for black hole entropy may
in some ways be universal\cite{frolov}: it  should not depend on the
form of the macroscopic gravitation theory, nor  on the details of the 
underlying
microscopic quantum theory.
This is also emphasized by recent work of Wald\cite{wald},
who showed that  black hole thermodynamics
and entropy are generic features of diffeomorphism invariant theories
with a curvature term in the action. 
It is therefore important
to examine black hole thermodynamics and statistical mechanics using
a variety of methods in as many different theories as possible. Identifying
model independent features might provide clues about the geometrical
source of black hole entropy.
\par
The purpose of the present paper is to examine the
thermodynamics of black holes in a large class of theories in
1+1 spacetime dimensions.
 Most derivations of the canonical partition function start 
from the Euclidean action\cite{whiting}. Recently, however, Louko and Whiting(LW) 
\cite{LW} applied the canonical formalism of Kuchar\cite{kuchar} to derive the Hamiltonian boundary terms for an eternal black hole inside a box of fixed radius and temperature. 
They restricted consideration to spacelike hypersurfaces that
ended  on the bifurcation point on the interior of an eternal
 black hole along a static slice. These boundary conditions are well
suited to the study of thermodynamics since they allow for analytic continuation to Euclidean spacetime. Remarkably, LW found
that these boundary conditions lead to a surface contribution to 
the Hamiltonian that was proportional to the entropy. They then used the resulting reduced Hamiltonian to derive the partition function
for a canonical ensemble of black holes. The same techniques have since been
 applied
to string inspired gravity\cite{SIG} by Bose {\it et al}\cite{bose1}, 
to Reissner-Nordstrom-anti de Sitter black holes\cite{louko1} and to
Lovelock black holes\cite{louko2}. 
In the following, we will show that the results of Louko and Whiting 
generalize easily to the case
of generic  vacuum 2-D dilaton gravity \cite{banks}.  In particular, the form of the reduced Hamiltonian and resulting partition function is essentially model independent\footnote{A preliminary form of this result, for slightly different
boundary conditions, was first reported in \cite{BK1}.}.
The classical thermodynamic entropy arises generically
as a surface term at the bifurcation point and the resulting Hamiltonian
looks more like a free energy\cite{bose2} than a Hamiltonian for
a pure quantum system. This result appears to lend
 support to Jacobson's conjecture\cite{jacobson} that the gravitational action
is best thought of as an effective theory describing the bulk properties of 
an as yet unknown underlying microscopic theory.
\par
 Instead of using the powerful, but technically involved formalism of Kuchar\cite{kuchar}, we apply the simpler techniques
applied extensively to generic dilaton gravity in a variety of recent papers\cite{domingo1}-\cite{BK2}. The key steps in this method are to
reparametrize the theory so as to make the generic action take a very
simple form, and then to rewrite the Hamiltonian constraint as the spatial divergence of the mass observable\cite{kuchar,thiemann,domingo1}. Our
analysis shows in a clear and completely general way how the thermodynamical entropy emerges from the canonical analysis. Moreover we are able to analyse the semi-classical limit of the resulting thermodynamic partition function generically to show the emergence
of the usual Bekenstein Hawking entropy/temperature relations. Finally we analyze in detail an 
interesting sub-class of models obtained by dimensional reduction of Einstein gravity in n+2
dimensions. We will show that the qualitative features of the partition
function are more or less independent of $n$: the specific heat is
always positive and there is for all $n$ a  phase 
transition from a high temperature, semi-classical phase to a low
temperature quantum phase dominated by zero mass (Planckian) black 
holes. This agrees with previous work\cite{whiting,LW}
for spherically symmetric gravity (which corresponds to the $n=2$ case).
For $n=2$ the transition is first order, but as $n$ increases the transition
``weakens'', i.e. resembles more closely a second order transition.
In the $n\to \infty$ limit, for which the theory resembles  string inspired dilaton gravity\cite{SIG}, the transition appears to be strictly second order.
\par
The paper is organized as follows: In Section 2, we review some important
features of generic dilaton gravity, including the action, space of solutions
and classical black hole thermodynamics.   Section 3 reviews the Hamiltonian analysis for the given
boundary conditions, and derives the necessary surface terms. The generic  partition function for a canonical ensemble of black holes in a box at fixed
temperature is derived and analyzed in the semi-classical approximation in
Section 4. Section 5 contains  results specific to  dimensionally reduced Einstein Gravity
in n+2 dimensions with n-dimensional spherical symmetry (SnG). 
Section 6 closes with conclusions and prospects for future work.

\section{Vacuum Generic Dilaton Gravity}
The most general action functional depending on the metric tensor
${\g}_{\mu\nu}$ and scalar field ${\ppi}$  in two spacetime
dimensions that  contains at most second derivatives of the fields can
be written\cite{banks}:
\bea
I[\g,\ppi]&=&{1\over2G}\int d^2x \sqrt{-\g}\left( \half
\g^{\alpha\beta}
\partial_\alpha \ppi \partial_\beta \ppi +{1\over
l^2} {V}(\ppi) +
D(\ppi)
R(\g)\right).
\label{eq: action 1}
\eea
where $R(\g)$ is the Ricci curvature scalar. The dilaton potential, $V(\ppi)$,
is an 
arbitrary function of the
dilaton  field. In the above, the fields $\ppi$ and
$\g_{\mu\nu}$ are taken to be dimensionless, as is the 2-D Newton constant, $G$. This requires
the inclusion of a coupling constant, $l$, of dimension length in the
potential term.
\par
If $D(\ppi)$ is a differentiable function of $\ppi$ such that $D(\ppi)\neq0$
and ${d D(\ppi)\over
d \ppi}\neq0$ for any admissable value of $\ppi$ then  the
kinetic term for the scalar field can be eliminated by means of the
(invertible) field
redefinition\cite{banks,domingo1}:
\bea
g_{\mu\nu} &= &\Omega^2(\ppi) \g_{\mu\nu} \\
\phi &=& D(\ppi)
\label{eq: field redefinitions}
\eea
where
\be
\Omega^2(\ppi) = \exp \left( \half\int {d\ppi \over (dD/d\ppi)}\right)
\ee
In terms of the new fields, the action \req{eq: action 1} takes the form:
\be
I = {1\over 2G}  \int d^2x  \sqrt{-g}\left(\phi
R(g)+{1\over
l^2}V(\phi)\right).
\label{eq: action 2}
\ee
where $V$ is defined as:
\bea
V(\phi) &=& {\Vbar(\ppi(\phi))\over \Omega^2(\ppi(\phi))}
\eea
This reparametrization accomplishes two things: it makes the
action and resulting Hamiltonian simpler by eliminating the kinetic
term for the scalar and it allows us to classify all possible theories
in terms of a single function of the dilaton field, namely the dilaton
potential. For example, spherically symmetric 4-D Einstein gravity\cite{SSG} is
described by a dilaton gravity theory with $V\propto 1/\sqrt{\phi}$, while
string inspired dilaton gravity\cite{SIG} corresponds to $V=constant$.
\par
In the following we consider the action in the form
\req{eq: action 2}.
The field equations are:
\be
R+{1\over l^2}{dV\over d\phi}=0
\label{eq: field1}
\ee
\be
\nabla_\mu\nabla_\nu \phi- {1\over 2l^2} g_{\mu\nu} V(\phi)=0
\label{eq: field2}
\ee
It follows directly from the above field equations that all 
solutions  have at least one  Killing
 vector given by\cite{DGK} 
\be
k^\mu = l\epsilon^{\mu\nu}\partial_\nu \phi/\sqrt{-g}
\label{eq: killing vector}
\ee
where the constant $l$ has been included to ensure that the
vector components are dimensionless. Note that the dilaton field is
also constant along this Killing vector.
\par
The most general solution to the field equations  in the
generic theory depends on one coordinate invariant parameter.
In a coordinate system adapted to the Killing vector (the analogue of 
Schwarzschild coordinates), the solution takes the form\cite{DGK,domingo2}:
\bea
ds^2&=&-(j(\phi) - 2GlM)dt^2 + {1\over (j(\phi) - 2GlM)} dx^2,
\nonumber\\
\phi &=& x/l,
\label{eq: general solution}
\eea
where $j(\phi) = \int^\phi_0 d\tilde{\phi} V(\tilde{\phi})$ and the parameter
$M$ will turn out to be the ADM mass. In these coordinates,
the Killing vector points along the time axis, from which one
can easily deduce that its
norm is
\be
|k|^2 = -l^2 |\nabla \phi |^2 = ( 2GlM - j(\phi)).
\label{eq: killing vector norm}
\ee
Given the above equation, it is clear that the general solution has an
apparent horizon at the surface $\phi= \phi_0 = constant$ for $\phi_0$
given by
\be
2GlM = j(\phi_0)
\ee
Whether or not this is an event horizon depends on the global properties
of the solution, which in turn depends on the form of the function $j(\phi)$.
We assume  that $j(0) =0$ and that
 $j(\phi)$ goes to infinity  monotonically as $\phi\to \infty$ so that 
$\phi_0$ is unique.
If in addition the surface $\phi=0$ is excluded from the manifold\footnote{
The effective Newton constant is infinite at $\phi=0$, and the field
redefinitions that we used to get to the simplified action in
general are singular there.} the resulting spacetime has precisely the same
global structure
as the radial part of  a Schwarzschild black hole. 

It is worth emphasizing
 that the global spacetime structure in these models derives
not only from the spacetime metric, but  from the structure of the 
dilaton field as well. This is not unreasonable given that one cannot
say {\it a priori} which metric is physical\footnote{i.e. to which metric
ordinary matter is minimally coupled.}, $g_{\mu\nu}$ or for example
$f(\phi) g_{\mu\nu}$. In fact in the present parametrization, the metric
corresponding to the
vacuum or matter free solution does not have all the properties normally 
 expected for an isolated Schwarzschild black
hole. 
The
Ricci scalar in \req{eq: field1}, which is the only independent curvature invariant 
in two spacetime dimensions,
 does not vanish in the asymptotic region $\phi\to\infty$, in general.
More importantly, the curvature does not depend directly on the mass parameter, and is non-vanishing when $M=0$. Thus the vacuum metric is not
 Minkowskian.
It is therefore
reassuring that  we can do a conformal reparametrization to a ``physical metric''
which does have most of the expected properties. In particular, define $\gphys$ by
\be
\gphys_{\mu\nu}= {1\over j(\phi)}g_{\mu\nu}
\label{eq: physical metric}
\ee
By transforming coordinates to $\tilde{x}$ such that
\be
d\tilde{x} = {d\phi \over j(\phi)}
\ee
one can see that the physical metric takes the usual Schwarzschild form:
\be
d\tilde{s}^2 = (1-2GlM/\tilde{j}(x))dt^2 - (1-2GlM/\tilde{j}(x))^{-1}d\tilde{x}^2
\ee
where $\tilde{j}(x)\equiv j(\phi(\tilde{x}))$. Note that this metric
approaches the Minkowski metric as $\phi\to \infty$ providing $j(\phi)\to\infty$ in this limit. Moreover,  the Ricci scalar
\bea
 R(\gphys)= \left[{GM\over l } \left(V' - {2 V^2\over jl}\right)\right]
\eea
does vanish when $M=0$. This metric is asymptotically flat and has 
a curvature singularity at $\phi=0$ for potentials of the form $V= k\phi^a$,
where $a<1$.

\par 
The thermodynamic properties of these black hole solutions are derived in
\cite{DGK}. Here we merely quote the results. The Bekenstein-Hawking entropy
for black holes in generic dilaton gravity is:
\be
S(M) = {2\pi\over G} \phi_0
\ee
where $\phi_0$ is the value of the dilaton at the horizon:
\be
\phi_0 = j^{-1}(2GMl)
\ee
The corresponding Hawking temperature is:
\be
T_H(M) = {V(\phi_0)\over 4\pi l}
\ee
It is important to note that the conformal 
reparametrizations expressed in \req{eq: field redefinitions} and
\req{eq: physical metric} do not affect the classical thermodynamics.

\section{Hamiltonian Analysis and Boundary Terms}
The Hamiltonian analysis for generic dilaton gravity has been presented in 
many works. Here we summarize the results, using the notation and conventions
of \cite{DK}. We start by decomposing the metric as follows:
\be
ds^2=e^{2\rho}\left[-{u}^2dt^2+\left(dx+{v}dt\right)
^2\right]
.\label{eq:adm}
\ee
where $x$ is a local coordinate for the spatial section $\Sigma$ and $\rho$,
${u}$ and ${v}$ are
functions of spacetime coordinates $(x,t)$. For convenience we work with the form of the
action in \req{eq: action 2}. The transition to the physical metric can be done by
a simple point canonical transformation that mixes $\rho$ and $\phi$, but
leaves the lapse and shift functions unchanged. In terms of the
 parametrization \req{eq:adm},
the action \req{eq: action 2} takes the form (up to surface terms):
\bea
I &=&\int dt\int^{\sigma_+}_{\sigma_-} dx[ {1\over G} ({\dot{\phi}\over {u}}
({v}\rho' + {v}' - \dot{\rho}) + {\phi'\over {u}} (  {u}{u}' - {v}{v}' + {v}
\dot{\rho}
 + {u}^2\rho' - {v}^2\rho')\nonumber\\
& &+\half{u} e^{2\rho} {V(\phi)\over l^2})
\label{eq: action 3}
\eea
In the above dots and primes denote differentiation with respect to time and
space, respectively, while $\sigma_+$ and $\sigma_-$ are the outer and inner spatial
boundaries.
The canonical momenta for the fields
 $\{\phi, \rho\}$ are:
\bea
\Pi_\phi & = &{1\over G{u}} ({v}\rho'+{v}'-\dot{\rho})
\label{eq: Pi phi}\\
\Pi_\rho &=& {1\over G{u}} (-\dot{\phi} +{v}\phi')
\eea
The momenta conjugate to ${u}$ and  ${v}$ vanish: these fields play the
role of Lagrange
multipliers that are needed to enforce the first class constraints associated
with diffeomorphism
of the classical action. A straightforward calculation
leads to the canonical
Hamiltonian (up to surface terms which will be discussed below):
\be
H_c= \int dx\left({v}\fcal+ {{u}\over 2G} \gcal\right)
\label{eq: canonical hamiltonian}
\ee
where
\bea
{\cal P}&=& \rho'\Pi_\rho +\phi'\Pi_\phi-\Pi'_\rho \sim 0
\label{eq: f constraint}\\
\gcal&=& 2\phi''-2\phi'\rho' -2G^2\Pi_\phi \Pi_\rho - e^{2\rho} {V(\phi)\over
l^2}\sim 0
\label{eq: g constraint}
\eea
are secondary constraints.
\par
Since 2-D dilaton gravity obeys a generalized Birkhoff theorem\cite{domingo2}
we expect there to be only one independent, diffeomorphism invariant
 physical observable, namely the mass of the black hole. The phase space, however,  must have even dimension, and it turns out that
there is  one physical pair of canonical variables. The first of 
these  is most easily derived by 
defining the following linear combination of constraints:
\bea
\tilde{\cal E} &:=&l e^{-2\rho}\left(-\phi' {\gcal} + G \Pi_\rho {\cal P}
  \right) \nonumber\\
&=& {\partial {\cal M}\over \partial x}
\label{eq: tildeE}
\eea
where:
\be
{\cal M}:= {l\over 2 G} \left(e^{-2\rho} ( G^2 \pi_\rho^2 - (\phi')^2)
  +{j(\phi)\over l^2}\right) = {1\over 2Gl}(|k|^2 + j(\phi))
\label{eq: define M}
\ee
Clearly $\cal M$ is a constant on the constraint surface. One can verify that
it commutes weakly with the constraints. As discussed in \cite{DGK}, the constant mode of $\cal M$ is a physical
observable, corresponding to the ADM mass of the solution\footnote{It also
corresponds to the Casimir invariant that characterizes solutions in the
Poisson sigma model approach\cite{strobl}.}.  We henceforth call $\cal M$ the mass observable,
 to distinguish it from the total Hamiltonian. Its conjugate momentum
\cite{DK}
\be
P_{\cal M} = -{G\over l}\int dx {e^{2\rho}\pi_\rho \over
   [(G\pi_\rho)^2 - (\phi')^2]}
\label{eq: conjugate to M}
\ee
is only invariant under diffeomorphisms that vanish at the boundaries of 
the system. Thus, the Hamiltonian analysis is consistent with the
generalized  Birkhoff
theorem.
It is possible, following
Kuchar\cite{kuchar} to do a canonical transformation to variables
such that 
$\{\cal M, P_{\cal M}\}$ are used as one pair of the phase space variables but this will not be
necessary for our purposes.
\par
In terms of the new constraint, the canonical Hamiltonian is:
\be
 H_c= \spatial
   \left(-\tilde{u}{\cal M}'+\tilde{v}{\cal P}\right) + H_+-H_-
\label{eq: canonical hamiltonian1}
\ee
where
\bea
\tilde{u} &=& {u e^{2\rho}\over l\phi'}\\
\tilde{v} &=& v+{uG\Pi_\rho\over \phi'}
\label{eq: tildev}
\eea
$H_+$ and $H_-$ are boundary terms
 determined by the requirement that the surface
terms in the variation of $H_c$ 
vanish for a given set of boundary conditions.
\par
For concreteness we will choose black hole boundary conditions considered
recently by
Louko and Whiting\cite{LW} in the case of spherically symmetric gravity. In
particular we consider the analogue of an eternal black hole in a box of fixed
constant radius. This requires keeping the value of the dilaton fixed
and time independent at ${u}_+$: $\phi({u}_+) = \phi_+$,
$\dot{\phi}_+=0$. The latter condition implies that $\tilde{v}_+=0$,
so that the only contribution to $H_+$ will come from the first
term in the Hamiltonian:
\be
\delta H_+({\cal M}) = \tilde{u}\delta {\cal M}|_{\sigma_+}
\label{eq: delta H+}
\ee
{}From $\tilde{v}_+=0$ and \req{eq: tildev} it follows that
$\Pi_\rho|_{\sigma_+} = - v\phi'/uG |_{\sigma_+}$. After substituting this equation into the definition
of the mass observable, a bit of algebra yields:
\be
\tilde{u}^2_+ = \left.{l^2 g_{tt}\over 2G {\cal M}l - j(\phi)}\right|_{\sigma_+}
\ee
Following \cite{LW} we fix the metric along the outer boundary,
$g_{tt}|_{\sigma_+} = constant= g^+_{tt}$, in which case \req{eq: delta H+} can be
integrated to yield:
\bea
H_+({\cal M})
&=&{Q^+ j(\phi_+) \over lG}\left(
1- \sqrt{1-{2G{\cal M}l\over j(\phi_+)}}
          \right)
\label{eq: H+}
\eea
where we have  chosen the integration constant so that $H_+\to0$ as ${\cal M}\to 0$. In the above $Q_+ = \sqrt{-g^+_{tt} /j(\phi_+)}$ gives the proper time at the box with
respect
to the physical metric $\tilde{g}_{\mu\nu} = g_{\mu\nu}/j(\phi)$ defined in
Section 2.
$H_+({\cal M})$ is the dilaton gravity analogue of the Brown-York
quasi-local energy\cite{brown}. As $\phi_+ \to \infty$, this goes to
the ADM mass
\be
H_+({\cal M})\to H_{ADM}= {\cal M}
\label{eq: ADM mass}
\ee
providing the physical metric is normalized to one at spatial infinity.

\par
Again following Louko and Whiting, we require 
the spatial slices to approach the
bifurcation point (i.e. $k^\mu =0$) along a static slice at $\sigma_-$. The conditions
that must be satisfied are\cite{LW}  $u_-=0$, ${v}_- =0$,
$\Pi_\rho(\sigma_-)=0$
and $\phi_-'=0$. With
these boundary conditions, the constraints imply that:
\be
\tilde{u}_- = {u'e^{2\rho}\over l \phi''} = {2 l  N_0\over V(\phi_-)}
\ee
where $N_0:= u'_-$ and we have used l'Hopital's rule to get the middle expression. The
final expression was obtained from the constraint \req{eq: g constraint} with
$\phi'=0$ and $\Pi_\rho=0$.
$N_0$ gives the rate of change of the unit normal to the constant $t$
surfaces at the bifurcation point\cite{LW}. For the on-shell Euclidean
black hole it is proportional to the Hawking temperature.
Using the fact that $|k|_-^2 =0$(cf \req{eq: killing vector norm}), one has
\be
\delta {\cal M}_- = {V(\phi_-)\over 2G l} \delta \phi_-
\ee
so that
\be
\tilde{u} \delta M_- = {N_0 \over Gl} \delta \phi_-
\ee
This can be integrated for fixed $N_0$, so that the total canonical
Hamiltonian is:
\be
H_c= \spatial \left(-\tilde{u}{\cal M'}+\tilde{v}{\cal P}\right)
 + H_+({\cal M}) - {N_0\over 2\pi} S({\cal M})
\label{eq: canonical Hamiltonian}
\ee
where we have defined
classical thermodynamic entropy\cite{DGK}:
\be
S({\cal M})= {2\pi\over G}\phi_- = {2\pi\over G} j^{-1}(2Gl{\cal M})
\ee
\req{eq: canonical Hamiltonian} generalizes Eq.(5.2) of \cite{LW} to the case
of black holes in generic dilaton gravity.
%%%%%%%%%%%%%%%%%%%%%%%%%%%%%%%%%%%%%%%%%%%
\section{Quantum Partition Function}

Generic dilaton gravity in 1+1 dimensions can be quantized exactly\cite{BK1,domingo1,strobl} by 
imposing the Hamiltonian constraints as operator constraints
on physical states. One can in fact find exact physical
 eigenstates of the mass observable\cite{domingo1,BK1,domingo3}. Alternatively,
one can  reduce the
phase space to the physical degrees of freedom at the classical level and then
quantize. Here we will follow the program of Louko and Whiting\cite{LW} and
do the latter. On the constraint surface $\cal M$ is 
independent of the spatial coordinates. The physical phase space 
consists of the mass observable ${\cal M}|_{\hbox{phys}}$ and 
its canonical conjugate.
It is convenient to introduce a dimensionless mass parameter:
\be
M := {\cal M}l
\ee
and dimensionless box size 
\be
B := j(\phi_+)/G
\ee
In terms of these, the  reduced action is simply:
\be
I = \int dt (P_M \dot{M} - H(M;{B},N_0))
\ee
withHamiltonian:
\be
 H(M;{B},N_0))=E({ M}, B) - {N_0\over 2\pi} S({ M}) .
\ee
$E({ M}, B)$  is the quasi-local energy given by:
\be
E({ M}, B)= { j(\phi_+) \over lG}\left(1- \sqrt{1-{2G
{ M}\over j(\phi_+)}
         } \right) = {B\over l}\left(1- \sqrt{1-{2M\over B}}\right)
\ee
while $S(M)$ classical entropy defined in the previous section:
\be
S(M) = {2\pi\over G} j^{-1}(2GM) \,\, .
\ee
 Note that we
have set $Q_+ = 1$ without loss of generality.
In addition to the dynamical variable
$M$, the Hamiltonian $ H(M;{B},N_0)$ depends on the  size, ${B}$, of the box in which the black hole  is placed and
the rate of change of the unit normal to constant $t$ surfaces at the
interior point, as given by $N_0$.

We would like to compute the quantum partition function:
\be
Z(\beta; {B}, N_0) = \hbox{Tr}\left[e^{-\beta \hat{H}(M;{B},N_0)}\right]
\ee
where the trace is over all physical mass eigenstates. This requires knowledge of the mass spectrum for the theory. However, 
since we are  effectively working in 
action-angle variables, we cannot gain any information about the
mass spectrum without making further assumptions. For example,
in reference \cite{BK2} it was shown that
by assuming periodicity of  the angle variable $P_M$, it is possible to derive
 a discrete mass spectrum for
Euclidean black holes. If the period corresponded to the inverse Hawking temperature, as required to make the Euclidean soliton solutions regular, the
spectrum was given by:
\be
S(M)  = 2\pi n
\ee
for any positive integer $n$.  
Remarkably, this is the same spectrum as obtained via Dirac quantization of 
Euclidean black holes in the generic theory\cite{BK1}.
Arguments also exist for the quantization of mass
in the Lorentzian sector as well\cite{bekenstein2}-\cite{mazur}\footnote{An
extensive list of references is given in \cite{makela}.} 

In the following,
we assume that the spectrum of the mass operator is bounded below by
zero and above by $M_+= j(\phi_+)/2Gl= B/2$, so that the exterior boundary 
remains outside the horizon of the corresponding black hole. In this case
the partition function can formally be written as a sum over mass
eigenstates $|M>$:
\bea
Z(\beta; {B}, N_0)&=& \int^{B/2}_0 dM \mu(M) <M|e^{-\beta\hat{H}}|M>\\
& =& \int^{B/2}_0 dM \mu(M) <M|M> e^{-\beta E(M;B) + {\beta N_0\over 2\pi}
     S(M)}
\label{eq: part 1}
\eea
One  important question concerns
 the choice of measure $\mu(M)$. It should be  a smooth function of positive weight. The
specific form of $\mu(M)$  should not affect
the qualitative features of the partition function, which we expect to be
dominated by the exponentials. For simplicity, we therefore choose
\be
\mu(M) = 1
\ee
 A more rigorous derivation of the measure
should be possible within the Dirac quantization procedure of \cite{BK1},
for example. This is currently under investigation.
\par
The expression \req{eq: part 1} for the partition function 
is divergent because the states $|M>$ are not normalizable.
As first argued in \cite{LW} it is reasonable to regulate 
this expression by replacing
\be
<M|M> = \delta(0)
\ee
by the inverse of the volume of the configuration space
so that the regularized partition function is:
\be
Z(\beta; {B}, N_0)= {2\over B}
\int^{B/2}_0 dM e^{-\beta E(M;B) + {\beta N_0\over 2\pi}
     S(M)}
\ee
\par
The only remaining ambiguity in this expression is the choice of $N_0$. 
Although our calculation is Lorentzian, we allow ourselves to be motivated by consistency with the Euclidean path integral, and choose:
\be
N_0= {2\pi\over \beta}
\ee
which is the value needed to avoid the conical singularity in the Euclidean
sector for solutions periodic in time with period equal to the inverse
temperature $\beta$. It is worth remarking that in the present context this
choice seems somewhat artificial: it leads to a Hamiltonian in the Lorentzian
framework that is explicitly temperature dependent\footnote{GK is grateful to
Valeri Frolov for useful discussions on this issue.}. We will nonetheless
follow the ``traditional path'' and see that it gives rise to potentially
interesting physics.

Finally, our  assuptions have lead us to the following quantum partition function for generic dilaton gravity:
\be
Z(\beta; {B})=
{2\over B}\int^{B/2}_0 dM e^{S(M)}  e^{-\beta E(M;B) }
\label{eq: partition 1}
\ee
It is interesting that the classical thermodynamical entropy generically
enters the integral as the logarithm of an apparent degeneracy of states
with mass $M$. However it must be
remembered that this derivation does not explain the degeneracy via
microscopic states, and therefore does not solve the black hole entropy
problem. It merely re-expresses it in a Hamiltonian context, and more 
importantly, shows that it is a completely generic feature of the class of
theories considered.

We close this Section by examining the
partition function in the 
semi-classical approximation. To this end we write the partition function
as:
\be
Z(\beta, B) = {2\over B}\int^{B\over2}_0 dM e^{-I(M;\beta,B)}
\label{eq: partition 2}
\ee
where
\be
 I \equiv \beta E(M,B) - S(M)
\ee
It is clear that if a stable minimum $\overline{M}(\beta,B)$ of $I$ exists,
then the partition function can be approximated to first order in
the semi-classical approximation by:
\be
Z(\beta, B) \sim e^{-I(\overline{M},B)}
\ee
so that the free energy obtained from the partition function is:
\bea
{\cal F}&=&-T\ln(Z)\nonumber\\
  &=&I(\overline{M},B)/\beta = E(\overline{M},B)-TS(\overline{M})
\eea
The minimum $\overline{M}$ is found as usual at the 
extremum of the free energy:
\be
\left.{\partial {\cal F}\over \partial M}\right|_{\overline{M}}
= \left.{\partial E(M,B)\over \partial M}\right|_{\overline{M}}
-  T 
\left.{\partial S(M)\over \partial M}\right|_{\overline{M}}
=0
\ee
Using the fact that
\be
\left.{\partial E(M,B)\over \partial M}\right|_{\overline{M}}
 = {1\over l\sqrt{1-{2\overline{M}\over B}}}
\ee
and that 
\be
l\left.{\partial S(M)\over \partial M}\right|_{\overline{M}}
\equiv \beta_H(\overline{M})
\ee
is the asymptotic inverse Hawking temperature associated with a black
hole of mass $\overline M$,
we find:
\be
\beta = \sqrt{1-{2\overline{M}\over B}}\beta_H({\overline{M}})
\ee
Thus, in the semi-classical approximation, the inverse temperature
$\beta$ of the box is related to the Hawking temperature by the
usual blue shift factor.
Note that the blue shift is calculated relative to the physical
metric introduced in Section 2. This is because the Hamiltonian
was normalized with respect to this metric. 
\par
One can also verify that in the semi-classical approximation, the entropy $S_{GF}$ of the gravitational field
obtained from the partition function is equal to  the classical
thermodynamic entropy
$S(\overline{M})$:
\be
S_{GF} = \left(1-\beta{\partial\over \partial \beta}\right) \ln
   Z = S(\overline{M})
\ee

%%%%%%%%%%%%%%%%%%
\section{$S^n$ Spherical gravity}
We now consider the case of dimensionally reduced Einstein gravity
in $n+2$ dimensions with $S^n$ spherical symmetry(SnG). Higher dimensional 
black holes of this form were analyzed in detail at the classical level
in \cite{meyers}. For $n=2$ this theory
corresponds precisely to spherically symmetric gravity,  while
in th $n=\infty$ limit it goes over to the String Inspired Dilaton
gravity model[SIG]. We start with the Einstein-Hilbert action in $n+2$
spacetime dimensions:
\be
I^{(n+2)}_{EH} = {1\over 16 \pi G^{(n+2)}} \int d^{n+2}x \sqrt{-g^{(n+2)}}
   R^{(n+2)}
\ee
and impose spherical symmetry via the  {\it ansatz}:
\be
ds_{(n+2)}^2 = {\g}_{\alpha\beta}(x,t)dx^{\alpha}dx^{\beta} +
  { r^2(x,t)} d\Omega^{(n)}
\ee
In the above, $x$ is a radial coordinate, $\{x^\alpha = x,t\}$ and $
\Omega^{(n)}$ is the volume 
form on the unit n-sphere and  $r(x,t)$ is 
the invariant radius of an n-sphere at $(x,t)$.

The dimensionally
reduced action takes the form:
\be
I^{(n+2)}_{EH}= {{\cal V}^{(n)} \over 16\pi G^{(n+2)}} \int d^2x
    \sqrt{-{\g}} {r}^n \left( R(\g) + {n(n-1) \over {r^2}}
  + n(n-1) \left|\nabla{\psi}\over{\psi}\right|^2 \right)
\label{eq: reduced sng}
\ee
where the volume of the unit $n$-sphere is:
\bea
{{\cal V}^{(n)}}&=& \int \Omega^{(n)}=  {2\pi^{(n+1)/2}\over \Gamma({1\over 2}
  (n+1))}
\eea
\req{eq: reduced sng} is of the form of a generic dilaton gravity theory in 1+1 dimensions. 
We now define a new, dimensionless
scalar $\ppi$:
\be
\ppi(x,t) = \left({r\over l}\right)^{n\over2}
\ee
 $l$ is an arbitrary constant with dimension length, which we take
without loss of 
generality to be the Planck length in $n+2$ dimensions:
\be
l^n\equiv G^{(n+2)}
\ee
In terms of $\ppi$, the action \req{eq: reduced sng} takes the form \req{eq: action 1} providing we make the following identifications:
\bea
{1\over 2G}&=&{8(n-1){\cal V}^{(n)} \over 16\pi n}
\label{eq: 2G}\\
D(\ppi) &=& {n\over 8 (n-1)} \ppi^2\\
\Vbar(\ppi) &=&  {n^2\over 8} \ppi^{(2n-4)/n}
\eea
We can now put the action in the canonical form \req{eq: action 2}
by
the  conformal reparametrization 
\req{eq: field redefinitions} with
\be
\phi = D(\ppi) = {n\over 8 (n-1)} \ppi^2
\ee
and
\be
\Omega^2(\ppi) = C\ppi^{2(n-1)\over n}
\label{eq: omega}
\ee
where $C$ is a constant of integration. The transformed
dilaton potential is:
\bea
V(\phi)&=& {n(n-1)\over C}\left({n\over 8(n-1)}\right)^{n+1\over n}\phi^{-1/n}
\label{eq: sng potential}
\eea
Note that the dilaton field can be expressed in terms of the invariant 
radius $r$ as follows:
\be
\phi = {n\over 8 (n-1)} \left({r\over l}\right)^n
\label{eq: invariant r}
\ee

In order to determine the constant $C$ in \req{eq: omega}, recall that the
ADM mass \req{eq: ADM mass} was derived by normalizing the time
component of the physical metric $\tilde{g}_{\mu\nu}=g_{\mu\nu}/j(\phi)$ to be unity at spatial infinity. In the present context, we would like 
the physical metric to coincide with the projection $\g_{\mu\nu}=\Omega^{-2}g_{\mu\nu}$ of the
higher dimensional metric. Clearly, this will be true
 if $\Omega^2 = j(\phi)$, which requires:
\be
C = {n^2\over 8 (n-1)}
\ee
With this choice,  the  dilaton potential is:
\be
V(\phi)= (n-1)\left({n\over 8(n-1)}\right)^{1/n}\phi^{-1/n}
\ee
\par

We can now apply the results of generic dilaton gravity to investigate the
thermodynamics of black holes in SnG. In terms of the dimensionless mass
parameter $M={\cal M}l$, the entropy is:
\bea
S(M) &=& {2\pi\over G} j^{-1}(2GM)\\ \nonumber
&=& {2\pi\over G} \left({8(n-1)\over n}\right)^{1\over(n-1)}
   \left({2GM \over n}\right)^{n\over (n-1)}\\ \nonumber
&=& 4\pi \left({16\pi\over n^n {\cal V}^{(n)}}\right)^{1/(n-1)}
   (M )^{n/(n-1)}\\  \nonumber
&=& c(n) M^{n\over n-1}
\label{eq: sng entropy 1}
\eea
where we have used \req{eq: 2G} and defined
\be
c(n) = 4\pi \left({16\pi\over n^n {\cal V}^{(n)}}\right)^{1/(n-1)} 
\ee
We note for future reference that $c(n) \to 4\pi/\sqrt{\pi n}\to 0$ as $n\to\infty$.
\par
If we express \req{eq: sng entropy 1} in terms of
 the invariant radius of the horizon(\req{eq: invariant r}), we get the expected result that the entropy is one quarter of the area of the
horizon expressed in Planck units\cite{meyers}:
\bea
 S({ M}) =  {1\over 4}{r_0^n {\cal V}^{(n)}\over G^{(n+2)}}
\label{eq: sng entropy 2}
\eea
A straightforward calculation reveals  that the  Hawking temperature for black holes in SnG is:
\bea
T_H({M})
 = {(n-1)\over 4  n c(n) l} M^{-1\over n-1}
\eea
For $n=2$ the above expressions give $S=
4\pi G^{(4)}{\cal M}^2$ and $T_H({\cal M}) = 1/8\pi G^{(4)}{\cal M}$ which are the
correct entropy and temperature for a Schwarzschild black hole. 
As $n\to \infty$, $S\to 4\pi {\cal M}l/\sqrt{\pi n}$, 
while
$T_H\to \sqrt{\pi n}/4\pi l$. These are the
entropy and temperature of black holes in String Inspired Gravity, up to a factor of $\sqrt{\pi n}$.

We will now examine in detail the partition function  \req{eq: partition 2}.
In the present case, it takes the form:
\be
Z(\beta, B) = {2\over B}\int^{B\over2}_0 dM e^{-I(M;\beta,B)}
\ee
where the ``action'' $I$ is:
\be
I = {\tilde\beta} B\left(1-\sqrt{1- {2M\over B}}\right) - c(n) ( M)^{n/n-1}
\label{eq: action n}
\ee
where ${\tilde\beta}= \beta/l$ is a dimensionless inverse temperature.
First
of all we note that this partition function  describes a thermodynamically stable system (i.e positive specific
heat) for all finite $n$. Moreover there exists an interesting phase structure.
Numerical plots of
 the logarithm of the partition function as a function of
$B$ and $\tilde \beta$ for $n=2$, $n=3$ and $n=9$ are presented 
 in Figs 1a), 1b) and 1c),
respectively. For $n=2$ there is a ``kink'' in the partition function that 
signals a phase transition from the semi-classical region at high
temperature (low $\beta$) to a quantum
phase consisting of a gas of microscopic black holes at low temperature\footnote{The flattened region at the top of each figure is a consequence of the
graphic presentation.}. This
was first described by York and Whiting for spherically symmetric gravity
\cite{whiting}. The transition appears to be strongly first order for large
box size, but weakens as $B$ decreases. In fact, at very low $B$ there appears
to be a vapour phase, which is neither pure semi-classical nor pure quantum.
As the dimension $n$ is increased, the phase transition appears to weaken
for all $B$. These qualitative features can also be deduced 
by examining $I$. 
This is plotted in Figs 2a), 2b) and 2c) for $n=2,3,9$ and  fixed box size.
At high temperatures the partition function is clearly dominated by 
a non-zero minimum, $\overline M$ of $I$, but the system 
goes through a first order phase transition as the temperature is decreased.
However, as $n$ increases, the transition is weakened: at the critical temperature the value of the mass at the minimum moves towards zero, and the height
of the potential barrier decreases. Remarkably, it is possible to solve
analytically for the critical temperature as a function of the mass $\overline M$ at the minimum. In the semi-classical phase, $\overline M$ approximates the
average black hole mass, and is obtained from:
\be
\left.{\partial I\over \partial M}\right|_{\overline M} = 0
\ee
which yields:
\be
\tilde{\beta}=  
   \sqrt{1- 2{\overline M}/B}{n c(n) \over n-1} {\overline M}^{1\over n-1}
\label{eq: minimum}
\ee
so that, as indicated in the previous section, the temperature of the box
is the red-shifted Hawking temperature for a black hole of mass $\overline M$.
At the cricital point, we also have that
\be
I({\overline M}) = \tilde{\beta}  B( 1-\sqrt{1- 2{\overline M}/B})
   - c(n) {\overline M}^{n\over n-1}=0
\label{eq: critical}
\ee
 Solving \req{eq: minimum}  for $2{\overline M}/B$ and substituting into
\req{eq: critical}, we find that the (dimensionful)
critical temperature is related to the
mean mass at the transition point by:
\be
{\beta}_c = {n c(n) l\over n+1} {\overline M}^{1\over n-1}
\label{eq: critical 2}
\ee
This formula can be explicitly verified by examining Fig. 3, where
 we plot the free energy at the critical point for $n=3$ for
a variety of box sizes.  Note that as $n\to\infty$, the critical temperature
goes to the constant $c(n)l$ and is independent of the box size and mean mass.

By substituting \req{eq: critical 2} into \req{eq: critical} it 
is possible to solve for the mean mass at the critical point
 as a function of the box size. It
is remarkably simple:
\be
{\overline M} = {2n\over (n+1)^2} B
\ee
This expression supports our claim that as $n$ increases, the first order transition becomes weaker: the mean mass at the critical point goes to zero. 
 
 Finally we note that, for all finite $n$ the partition function diverges
as the box size is taken to infinity. That is, as $B\to \infty$,
\be
I\to \tilde\beta M - c(n)M^{n/n-1}
\ee
Clearly, the second term dominates for all finite $n$ for any temperature as $M\to \infty$,
so that the integral will diverge, as claimed. 
\par

\section{Conclusions}

We have examined in some detail the Hamiltonian thermodynamics of black holes
in generic 2-D dilaton gravity. As the SnG example shows, these models can in some cases be
thought of as truncated higher dimensional theories, and are therefore not
simply ``toy models''. One important result of our analysis
is  that the classical thermodynamic
entropy generally contributes to the boundary term at the bifurcation point
in such a way as to make the Hamiltonian look like a free energy\cite{bose2}.
The consequences of this were examined in detail for the partition function
of SnG.
\par
The main puzzle in black hole thermodynamics is how the most simple gravitational system (i.e. a spherically symmetric black hole) can have sufficient
complexity to account for the Beckenstein-Hawking entropy. This puzzle is highlighted by our analysis, which shows explicitly in a Hamiltonian context
how an apparent ``degeneracy" of states generically 
arises in the partition function, despite
the fact that the mass eigenstates are non-degenerate.
This interesting behaviour can be interpreted as  support for
 the conjecture by Jacobson\cite{jacobson,frolov} that the gravitational action should be thought of as a bulk action induced by the interactions of microscopic quantum fields. In the context of such an interpretation, it is natural 
that the Hamiltonian for the gravitational field encodes the degeneracy of
the states in the fundamental microscopic theory. 

\section{Acknowledgements}
\par
GK is grateful to A. Barvinsky, J. Gegenberg and J. Louko and
B. Whiting for helpful
discussions.   This work
was supported in part by the Natural Sciences and Engineering
Research
Council of Canada. GK is grateful to the  CTP at MIT for
its hospitality during the initial stages of this work. 
  \par

%\bibitem{teitelboim} C. Teitelboim, ``Statistical Thermodynamics of
%a Black Hole in Terms of Surface Fields'', hep-th 9510180, 1995.
%\bibitem{gk} J. Gegenberg and G. Kunstatter, Phys. Rev. {\bf D47},
%R4192 (1993).
\clearpage
\centerline{
\epsfbox{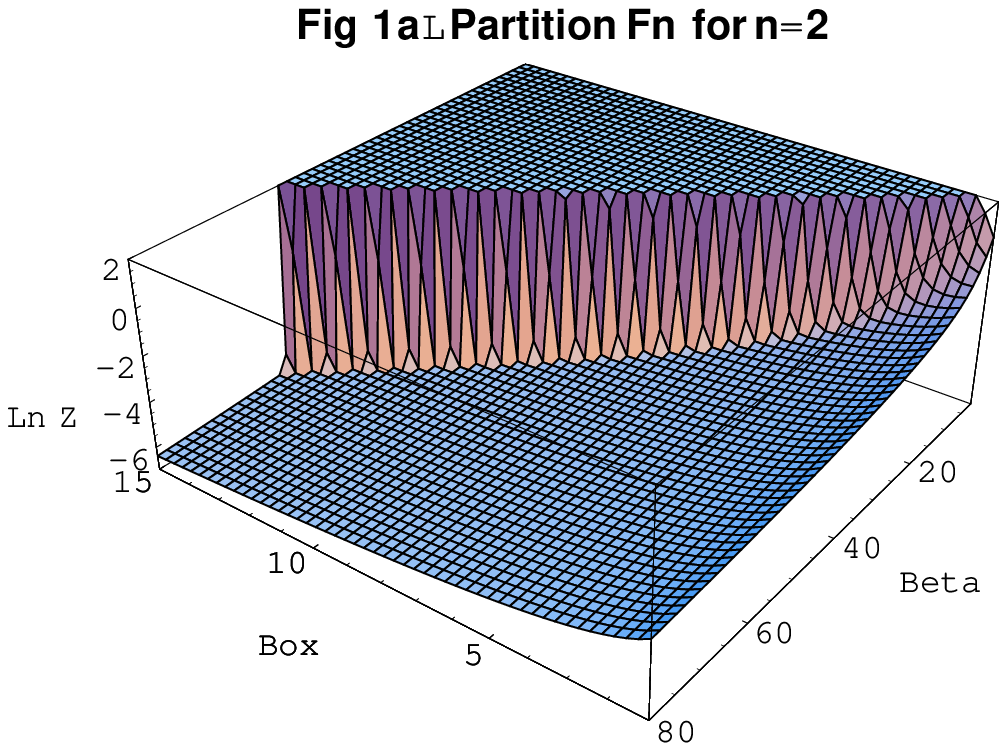}
}
\clearpage
\centerline{
\epsfbox{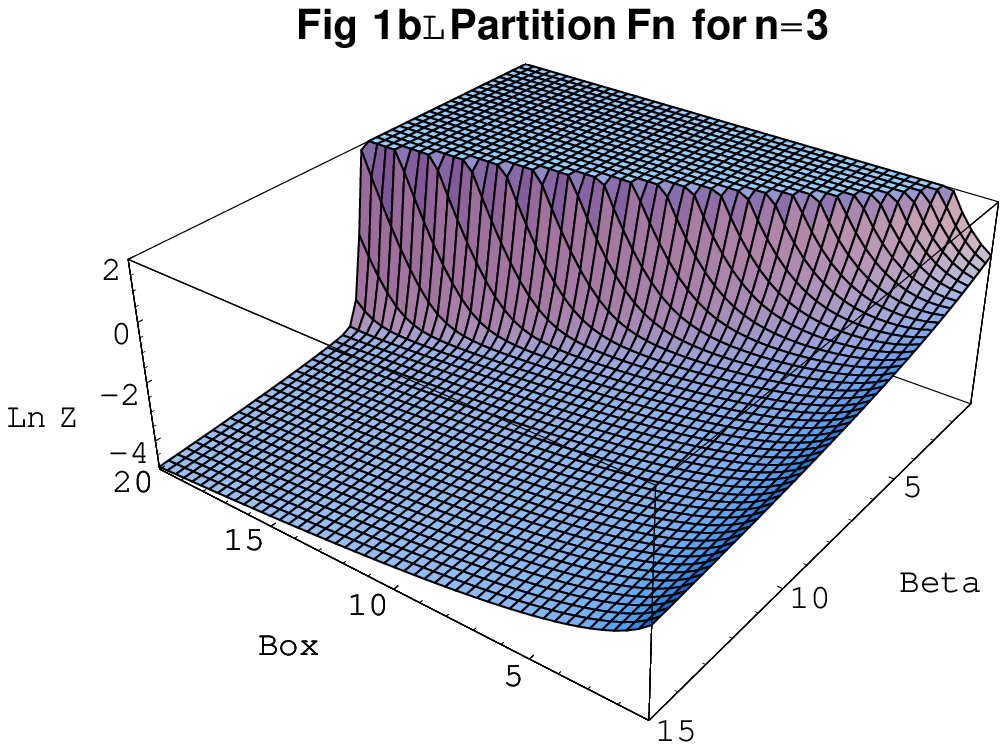}
}
\clearpage

\centerline{
\epsfbox{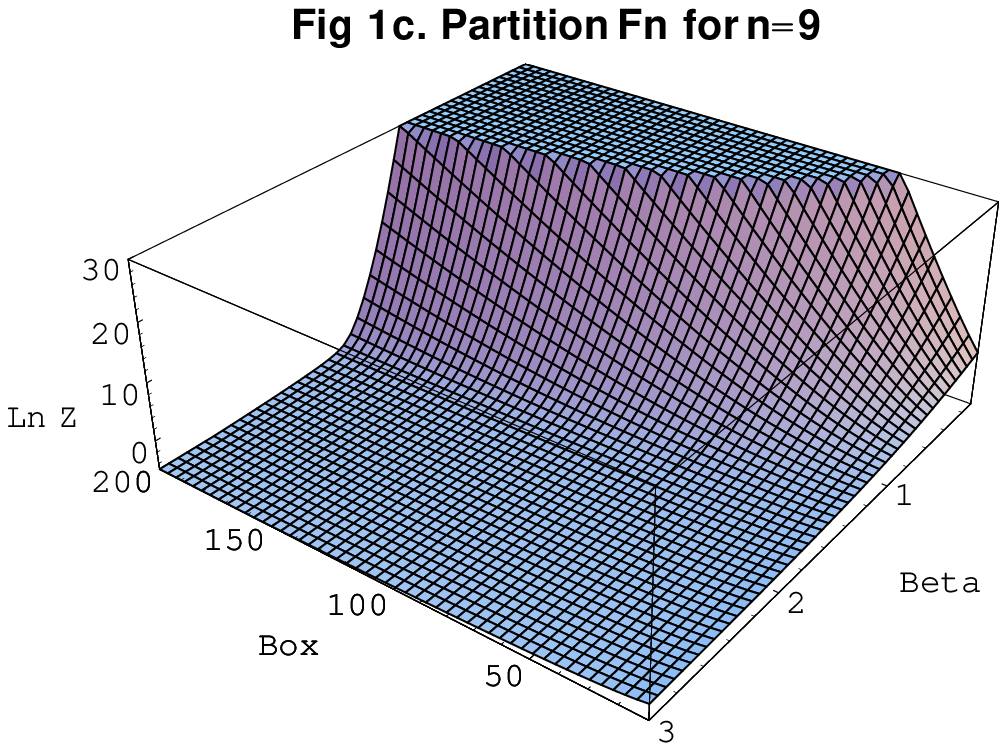}
}
\clearpage
\centerline{
\epsfxsize=7 in
\epsfbox{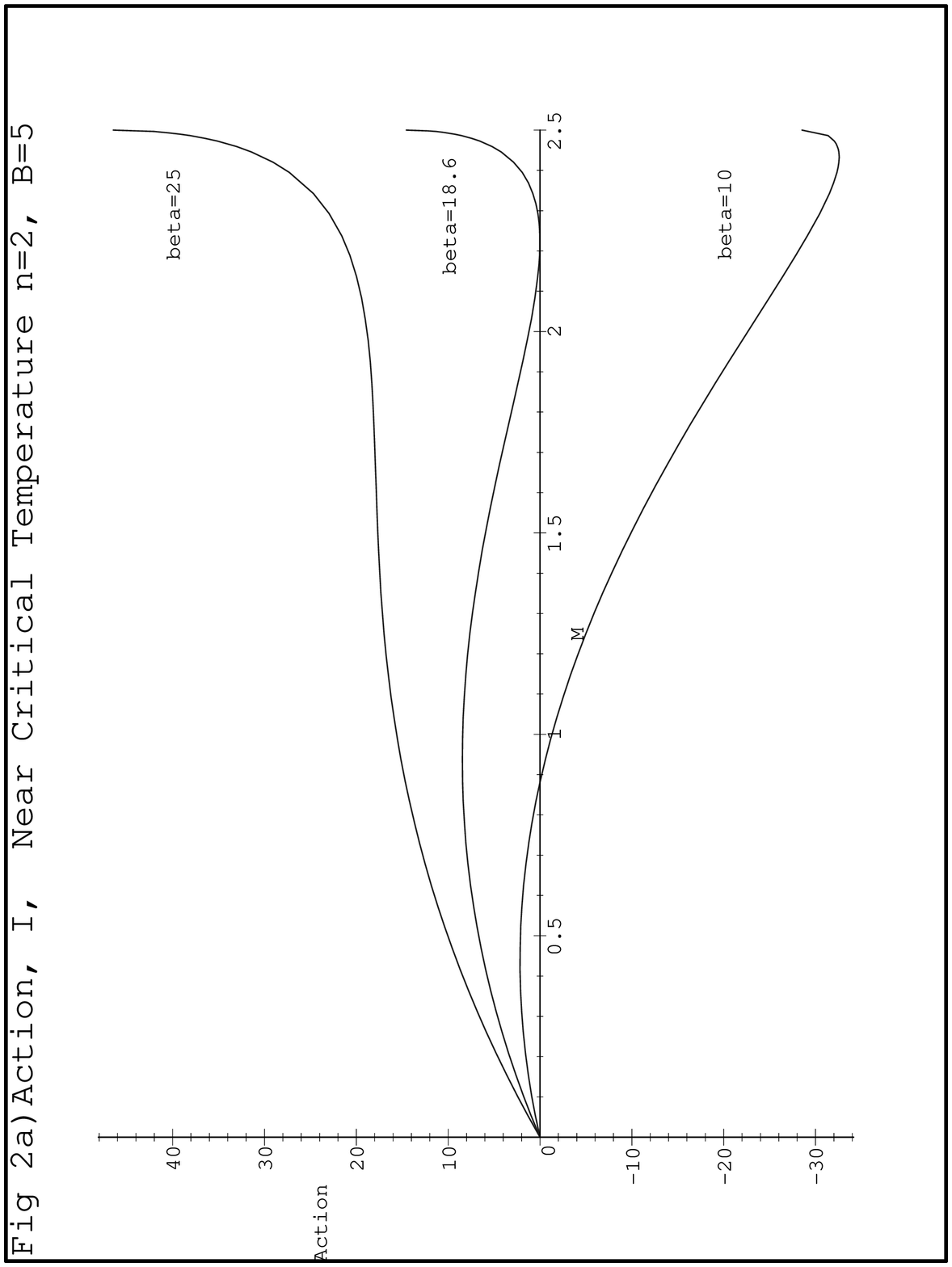}
}

\centerline{
\epsfxsize=7 in
\epsfbox{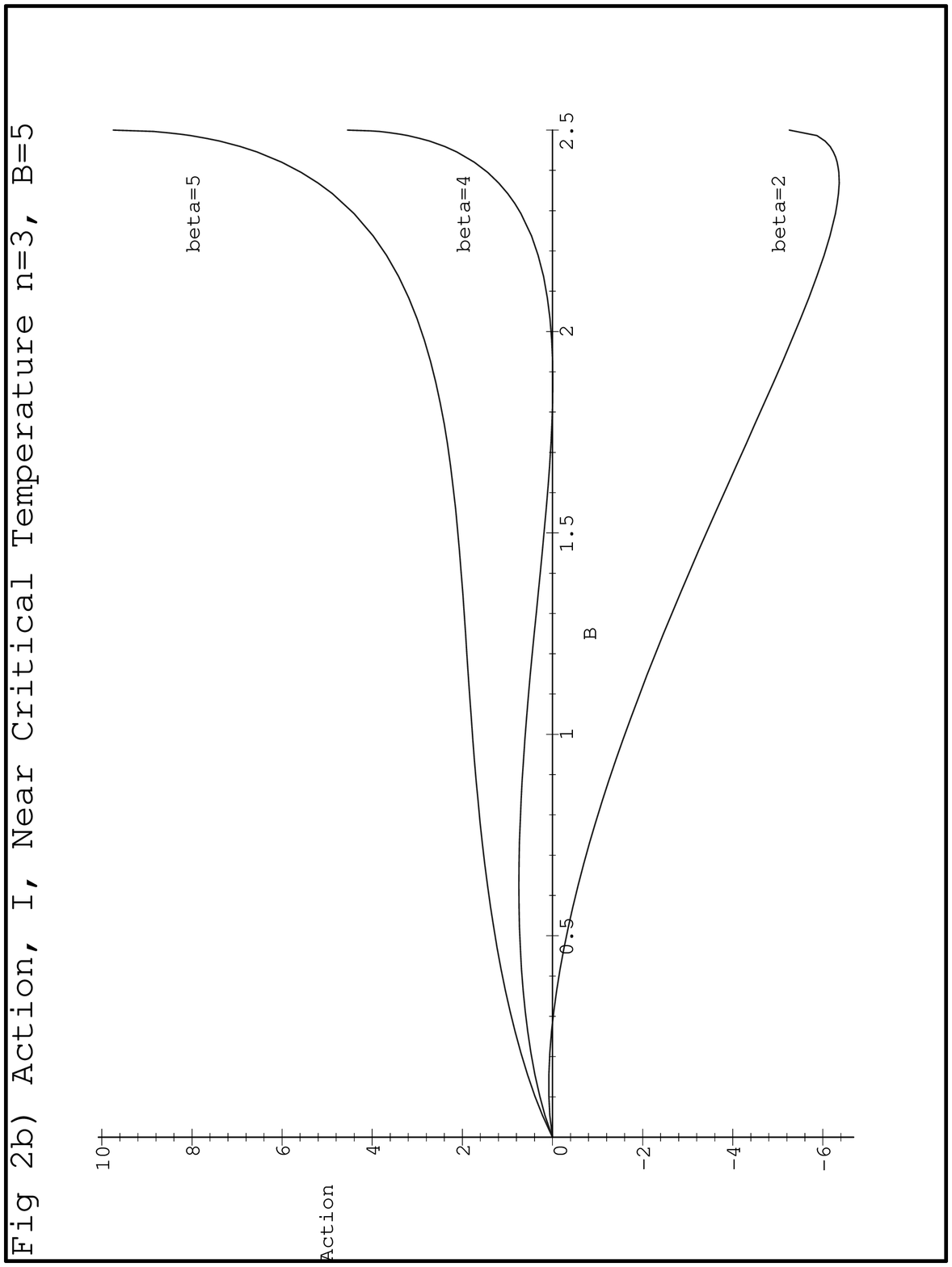}
}

\centerline{
\epsfxsize=7 in
\epsfbox{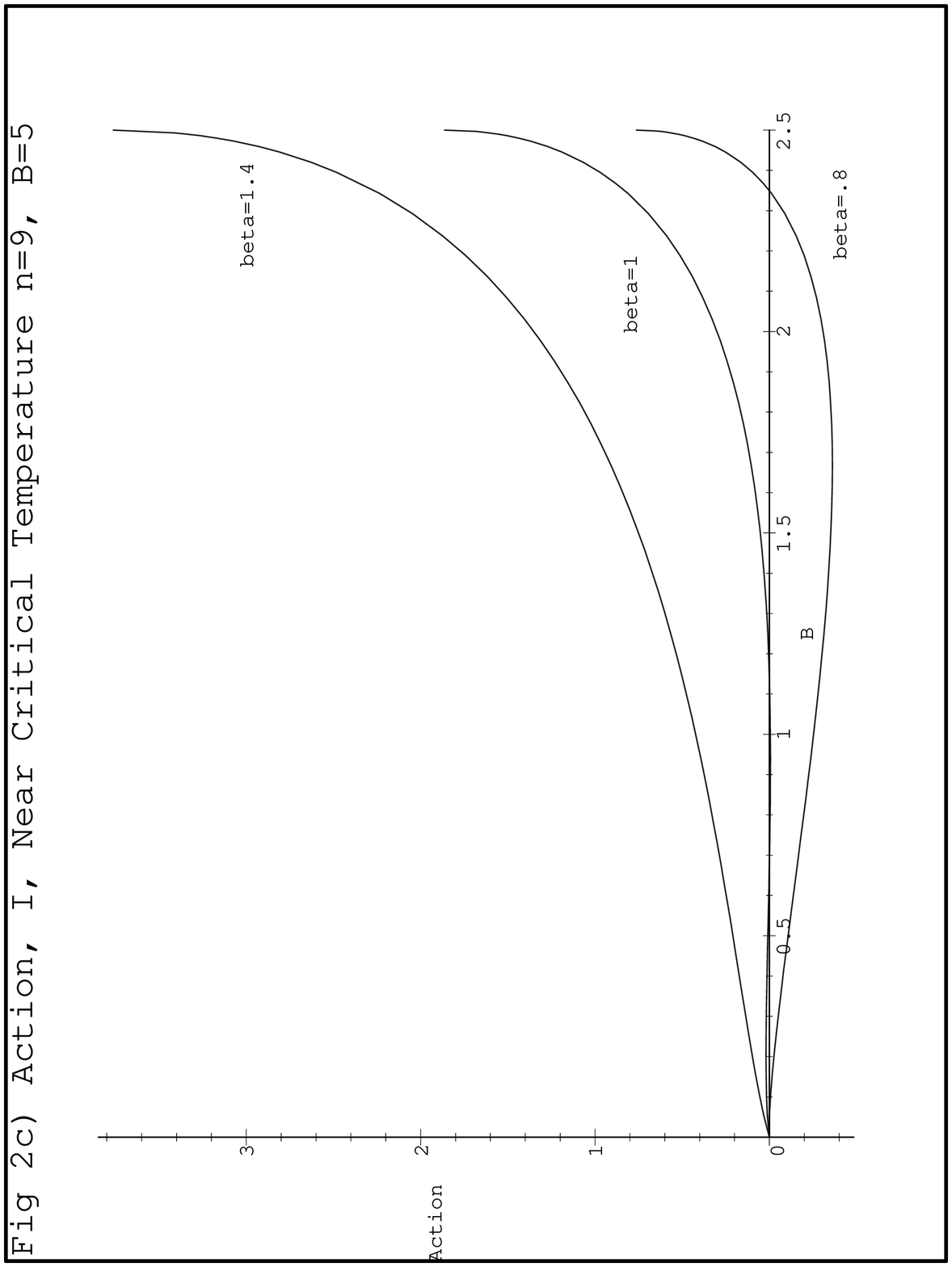}
}

\centerline{
\epsfxsize=7 in
\epsfbox{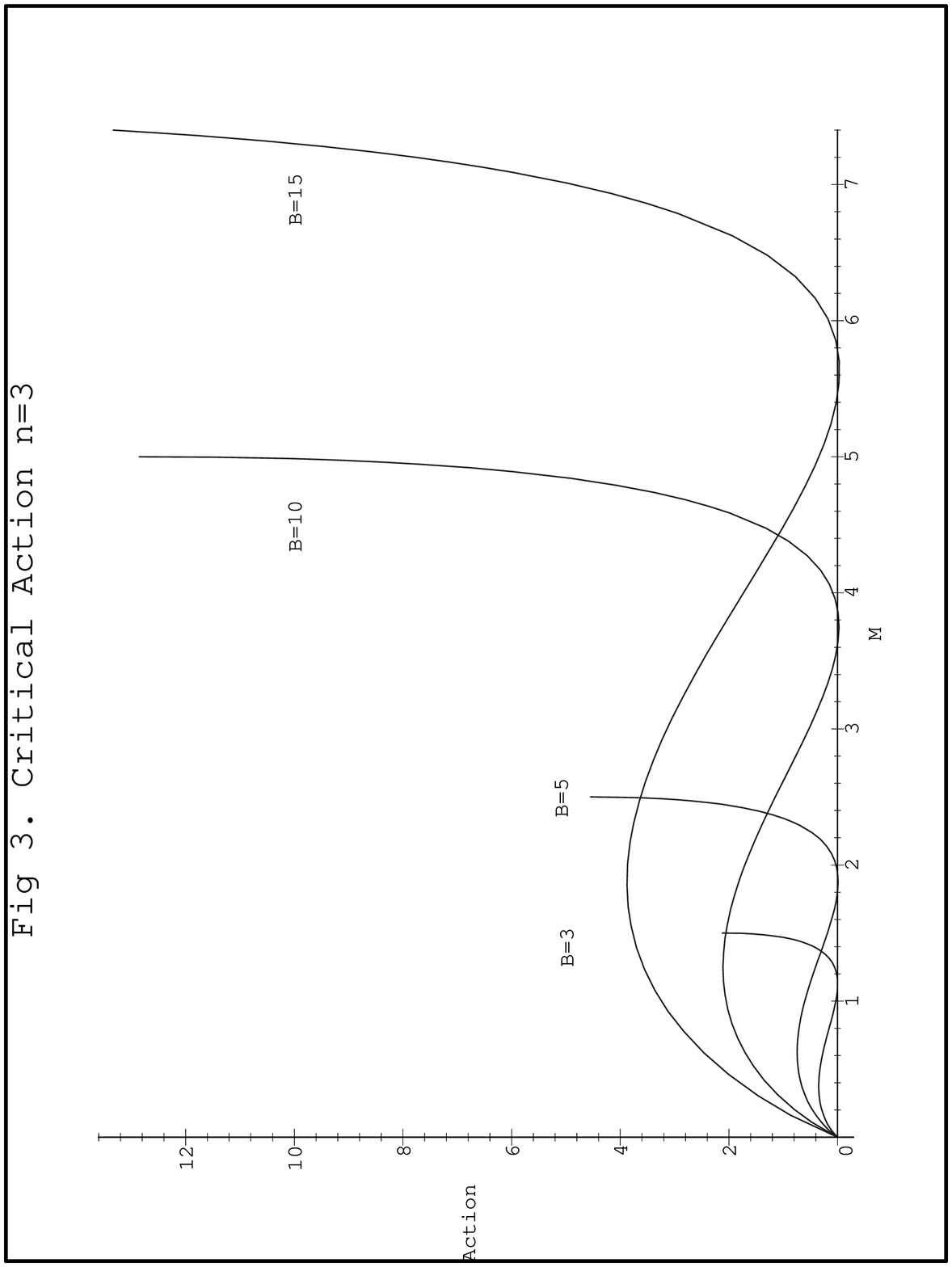}
}

\end{document}